\documentclass[preprint,showpacs,preprintnumbers,showkeys,a4paper]{revtex4}
\usepackage{graphicx,color}
\usepackage{bm}

\newlength{\imagewidth} 
\newlength{\halfimagewidth} 
\begin{document}
\title{The  energy landscape networks of spin-glasses}

\author{Hamid \surname{Seyed-allaei}}
\affiliation{Department of Physics, Sharif University of Technology (SUT),  P.O. Box 11365-9161, Tehran, Iran}
\author{Hamed \surname{Seyed-allaei}}
\affiliation{Department of Physics, Shahid Beheshti University (SBU), Evin, Tehran 1983963113, Iran}
\affiliation{School of Physics, Institute for Studies in Theoretical Physics and Mathematics (IPM),  P.O.Box: 19395-5531, Tehran, Iran}
\author{Mohammad Reza \surname{Ejtehadi}}
\affiliation{Department of Physics, Sharif University of Technology (SUT),  P.O.Box 11365-9161, Tehran, Iran}

\preprint{IPM/P-2007/064}
\date{\today}

\begin{abstract}
We have studied the topology of the energy landscape of a spin-glass model and the effect of frustration on
it by looking at the connectivity and disconnectivity graphs of the inherent structure. The connectivity network
shows the adjacency of energy minima whereas the disconnectivity network tells us about the heights of the
energy barriers. Both graphs are constructed by the exact enumeration of a two-dimensional square lattice of a
frustrated spin glass with nearest-neighbor interactions up to the size of 27 spins. The enumeration of the
energy-landscape minima as well as the analytical mean-field approximation show that these minima have a
Gaussian distribution, and the connectivity graph has a log-Weibull degree distribution of shape $\kappa=8.22$ and
scale $\lambda=4.84$. To study the effect of frustration on these results, we introduce an unfrustrated spin-glass model
and demonstrate that the degree distribution of its connectivity graph shows a power-law behavior with the
$-3.46$ exponent, which is similar to the behavior of proteins and Lennard-Jones clusters in its power-law form.
\end{abstract}

\pacs{89.75.Hc, 05.50.+q, 05.10.-a}
\keywords{Energy landscape, Spin-glass, Inherent Structure, Connectivity network, Disconnectivity network, Exact Enumeration}

\maketitle

\section{Introduction}
The dynamic and structure of complex systems can be approached via their energy landscape\cite{2004enla.book.....W, CharlesL.BrooksIII07272001} some examples are protein folding problem\cite{bryngelson1995raf,onuchic1997tpf}, Lennard-Jones liquids\cite{PhysRevA.25.978, JPhysCondensMatt.18.6507} and spin glasses\cite{amoruso2006deb, PhysRevLett.86.3148, PhysRevE.73.036110, 2002PhyA..315..302K}. We can however construct the energy landscape only for systems of small size, because in general the phase-space volume grows exponentially as the system's degrees of freedom increase. To tackle the restriction regarding the size of the phase space, we can sustain the key information of the landscape, such as the energy minima (inherent structure) and transition states between minima, and encode them into two kind of complex networks\cite{newman_siam} that are known as connectivity and disconnectivity graphs. 
In a connectivity network, each node represents a local minimum energy of the system while the links are transition pathways connecting two neighboring minima. In a disconnectivity network, each node represents a super basin with an energy of $E$. The super basin of energy $E$ consists of at least two smaller super basins, which are separated by an energy barrier of the absolute height $E$. In this way we can construct a hierarchical tree in which each super basin is connected to its subset super basins. In such a tree, the root is the largest super basin that includes all the states, and leafs are the local energy minima. The disconnectivity graph was first introduced by Becker and Karplus to study peptide structure and kinetics\cite{1997JChPh.106.1495B}.

Complex networks have been frequently employed  to analyze the energy landscapes of different physical  systems\cite{doye_PhysRevLett.88.238701,kabakcciouglu2005sfn, amaral2001swn, amaral2000csw,tree.spinglass.2003}. This method also has been utilized in the consideration of the energy landscape of the system of Lenard-Jones particles\cite{doye_PhysRevLett.88.238701}, the protein chains\cite{rao2004pfn,gfeller2007ccn} or spin-glasses\cite{PhysRevE.60.3219}. 

Networks are usually categorized in terms of their degree distribution, i.e. the distribution of the number of links at each node in the network. A well studied kind of network in the literature is the scale-free network which has a power-law form for the degree distribution\cite{albert_barabasi}. The scale-free behavior has been observed  in various systems such as friendship, the Internet\cite{newman_siam}. The same also has been reported by Doye in the energy landscape of a cluster of Lenard-Jones particles\cite{doye_PhysRevLett.88.238701}, in short peptide chains by Rao and Caflisch\cite{rao2004pfn} and Gfeller \textit{et all}\cite{gfeller2007ccn}. Burda {\it et al} also reported a power-law tail only for a spin-glass model embedded in a random graph\cite{PhysRevE.73.036110}, in other cases they showed that a one-dimensional spin-glass exhibits a normal or log-normal distribution of nodes' degree. It is still unclear if frustrated systems with a very rough energy landscape are also a scale-free.

Spin-glasses in general are good prototype models to examine the rough energy landscape of frustrated systems. They represent a broad class of the complex systems. Many interesting problems are indeed analogous to spin-glasses, examples are the three satisfiability problem\cite{zecchina_PhysRevLett.76.3881} or the protein folding\cite{bryngelson1987sga}. 
Setting aside the advantage of the Derrida's random energy model (REM)\cite{rem_PhysRevLett.45.79} which simplifies the analytical studies of the spin-glass systems and, in part, defeat the purpose of such studies. 

This article is organized in the following order: In Sec.~\ref{sec:mean}, we obtain an analytical estimate of the probability density of energy minima and degrees of nodes. In Sec.~\ref{sec:enum}, we state the enumeration method which includes helical boundary condition, multiple spins method with the help of bit-wise operations (the most basic operators on a computer, e.g., AND and XOR)  and a painting algorithm to label minima and their surrounding basins. Afterwards, in Sec.~\ref{sec:res} we discuss the enumeration results, and finally, we compare our results with other energy landscape networks and discuss the most relevant sources of differences, i.e., the presence of frustration and the energy density function. The concluding remarks are presented in Sec.~\ref{sec:con}.


\section{Mean field calculation}
\label{sec:mean}
\begin{figure}
   \centering
   \includegraphics[width=\imagewidth]{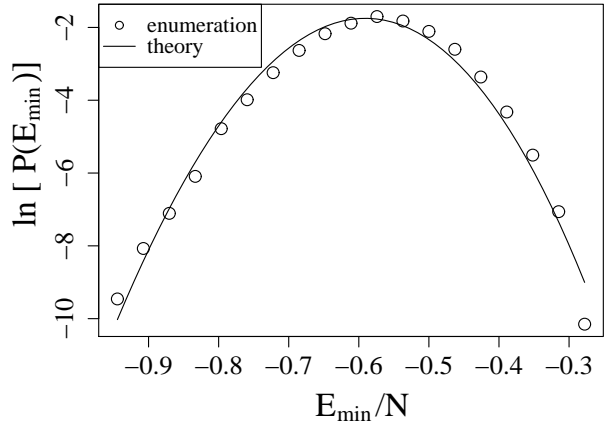}
   \caption{Comparison of the theoretical (line) and enumeration (disks) results. The line represents the Eq.~(\ref{eq:p_e}). The system size here is 27 spins and the PDF is averaged over 150 realizations.}
\label{fig:comparison}
\end{figure}
We consider a random Ising model on a 2D square lattice of $N$ sites. We present the spin in the $i^{th}$ site by $s_i$. Every spin is either up (+1) or down (-1). We represent each configuration of the system with a vector of spins denoted by $\mathbf{s} = (s_0, s_1, s_2, ..., s_i, ..., s_{N-1})$. The phase-space thus consists of $2^N$ such vectors.

We define a hamming  distance between two configurations of the system in the phase space, $\mathbf{s}$ and $\mathbf{s'}$, as the number of their mismatched elements $\sum (1-\delta_{s_i,s'_i})$. If the distance between two configurations is one --meaning that they differ only in one spin-- we call them adjacent neighbors.

We will consider a spin-glass-dynamics such that spins interact only with their adjacent neighbors. So, each spin interacts with $4$ of its neighbors. We consider further that the coupling constant of the interactions, $J_{ij}$, is either $+J$ or $-J$. Then the  Hamiltonian follows
\begin{equation}
H = \frac{1}{2} \sum_{<ij>} J_{ij} s_i s_j, \label{hamiltoni}
\end{equation}
where summation is over the adjacent neighbors. We can rewrite the Hamiltonian as $H = J (n^+ - n^-)$ where $n^-$ is the number of satisfied interactions while $n^+$ is the number of frustrated interactions. The sum of these two numbers is $n^+ + n^- = 2 N $.

Random energy model\cite{rem_PhysRevLett.45.79} assumes that the energy density function is Gaussian.
\begin{eqnarray}
P(E) = (N \pi J^2)^{-1/2} exp(-E^2/N J^2),
\end{eqnarray}
wherein the energy of randomly chosen  state of $\mathbf{s}$ is $E$.

To begin with, we would like to calculate the probability that a given configuration $\mathbf{s}$ is a local minimum. For doing so, we first review the properties of local minima. The energy of a local minimum by definition is lower than the energy of its neighbors.  It's energy,  therefore, increases when we flip any of its spins.  As a result,  for a local minimum configuration, just one of the four interactions of any site can be frustrated. In other words, there must not exist any frustrated pairs of interactions sharing one end (spin). Consequently,  the odds for a configuration to be a local minimum is the same as of a configuration wherein all frustrated interactions are distinct (isolated from each other).

The probability to have a frustrated interaction surrounded by satisfied interactions is $(\frac{n^-}{2 N})^6$, where $n^-$ is the total number of satisfied interactions. The exponent of six is due to the number of the interactions which are connected to the frustrated interaction. Since we have $n^+$  unsatisfied (frustrated) interactions, the probability of having all the unsatisfied interactions isolated is $(\frac{n^-}{2 N})^{6 n^+}$. But here we have over counted some of the satisfied interactions. The over counting happens when a satisfied interaction has two frustrated interactions at its ends. Such cases approximately occur  $n^- (\frac{3 n^+}{2 N})^2$ times; therefore, the final probability of having a local minimum with energy $E$ is $\approx (\frac{n^-}{2 N})^{6 n^+-n^- \frac{9}{4}(\frac{n^+}{N})^2}$. Recalling $2 N = n^+ + n^-$ and $\frac{E}{J} = n^+ - n^-$,  we restate the probability  in terms of $E$ and $N$. This leads to
\begin{equation}
\left (\frac{1}{2}-\frac{E}{4 N J}\right)^{\frac{3}{32} (4+E/N J) \left(10 + (E/N J\right)^2)},
\end{equation}
which  is the probability distribution function (PDF) of energy minima to be derived from:
\begin{widetext}
\begin{equation}
P(E) = \frac{1}{\sqrt{N \pi J^2}} \\  e^{- \frac{E^2}{N J^2}} \   \left ( \frac{1}{2}-\frac{E}{4 N J}\right )^{\frac{3}{32} (4+E/ N J)\left(10 + (E/ N J)^2\right)}.
\label{eq:p_e}
\end{equation}
\end{widetext}

Having obtained the PDF of energy minima, it is easy to estimate the degree distribution. Each minimum is down in a basin. The volume of a basin is proportional to $e^{-E/J}$. The number of neighbors of a basin is proportional to its volume. Therefore, we can say that $\ln k \sim -\frac{E}{J}$ where $k$ is the degree of the node. Thus, the probability distribution of the degrees is approximated by:
\begin{widetext}
\begin{equation}
P(k)   \sim  e^{- \frac{\ln(k)^2}{N}-\ln(k)} \ \ \left (\frac{1}{2}+\frac{\ln(k)}{4 N} \right )^{\frac{3}{32} \left (4-\ln(k)/N \right ) \left (10 + \ln(k)^2/N^2\right )} .
\label{eq:p_k}
\end{equation}
\end{widetext}
We shall shortly see that Eq.~(\ref{eq:p_k}) leads to a good approximation for the distribution. In order to further clarify the Eq.~(\ref{eq:p_e}) and Eq.~(\ref{eq:p_k}) we show that within some approximations they reduce to familiar expressions. 
For small values of $\frac{E}{N J}$ we can approximately write
\begin{equation}
\left ( \frac{1}{2}-\frac{E}{4 N J}\right )^{\frac{3}{32} (4+E/ N J)\left(10 + (E/ N J)^2\right)} \approx
 2^{-\frac{15}{4}} e^{- \frac{15 E}{16 N J}}\ ,
\end{equation}
whose insertion into Eq.~(\ref{eq:p_e}) leads to the Gaussian distribution of 
\begin{eqnarray}
P(E) \approx \frac{1}{2^{\frac{15}{4}} \sqrt{N \pi J^2}} \ \  e^{- \frac{E^2}{N J^2}- \frac{15 E}{16 N J}}. 
\label{eq:pdf_e_simple}
\end{eqnarray}
Knowing the relation between $E$ and $k$ ( $\ln k \sim -\frac{E}{J}$), we get a simplified form for the PDF of degrees:
\begin{eqnarray}
P(E) \sim \frac{1}{2^{\frac{15}{4}} \sqrt{N \pi J^2}} \ \ e^{- \frac{\ln(k)^2}{N}- (1 -\frac{-15}{16 N}) \ln(k)}.
\label{eq:pdf_k_simple}
\end{eqnarray}

\section{Enumeration and algorithms}
\label{sec:enum}
We are going to construct the network of adjacent minima of a spin glass with $N$ spins and a set of given coupling constants. We achieve this goal by first sweeping the energy landscape. We generate all of $2^N$ configurations before calculating the energy of each configuration. Having computed the energy of all configurations, we then find the local minima with the help of a painting algorithm. Afterwards, we construct the adjacent network of the local minima. By employing the multiple spins method, we save CPU time and memory.  This enables us to sweep the  phase-space corresponding to systems of up to 27 spins, with an ordinary PC (about 3GHz and 2GB). We utilize the helical boundary conditions to take advantage of the periodic boundary conditions, and furthermore, to be able to study any number of spins, even though it is not a complete square\cite{newman1999mcm}.

\begin{figure}
   \centering
   \includegraphics[width=\imagewidth]{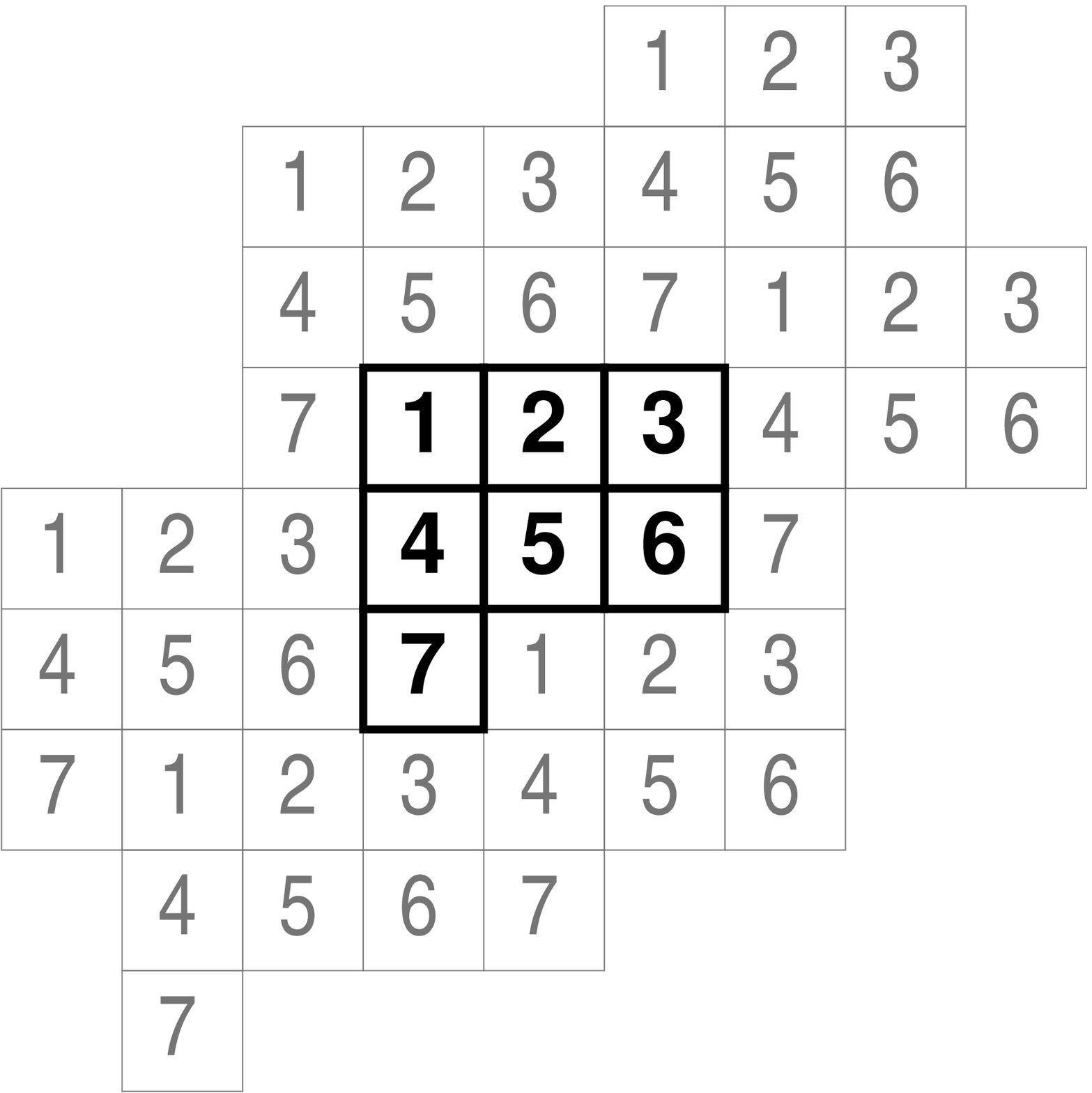}
   \caption{Illustrating the helical boundary condition. The lattice has 7 sites, 2 sites less than a complete square 9. Nevertheless, the plane is covered by lattices consistently.}
\label{fig:helical}
\end{figure}

Helical boundary condition resembles the periodic boundary condition but sets the last spin of a row to be the neighbor of the first spin of the next row (FIG.~\ref{fig:helical}). Helical boundary condition, avoiding the constraint of the square number of the lattice  sites, allows for an arbitrary number of spins, including the integer numbers which are not a complete square.

In the multiple spins method, each spin is represented by one single low-level bit of the computer memory: either 0, or 1. Therefore, a lattice configuration represented by $\mathbf{s}=(1,0,0,1,0,1,0)$ becomes a binary number $1001010$, which is $u=74$ in the decimal base.

We can also treat the coupling constants in the same way. The only difference is that the number of interactions are twice the number of spins.

Now we can use bit-wise operations, {\em exclusive or} (xor) and {\em bit rotation}, to calculate energy of a given state. First we calculate $(u\ \oplus\ R(u,1)) \oplus j_\_$, where $\oplus$ is xor operator and $R(u,n)$ rotates $u$ by $n$ bits. We count the number of ones in the binary form of the results which gives us the energy of the horizontal interactions. In the same way, we calculate the energy of vertical interactions. This time we have to rotate bits by the number of columns of the lattice $n_c$, $(u \oplus R(u,n_c)) \oplus j_|$. Again, we count the number of ones in the binary form of the result. The sum of these two counts gives us the total energy.

Since it is not very straightforward to deal with degenerate ground states in the painting algorithm, we used uniformly distributed coupling constants between $\pm 1$, to overcome degeneracy\footnote{The original multiple spin method applies only to coupling constants with values of $\pm1$. Nevertheless, it can be extended to our case, floating point constants. To do this, we need more than one set of coupling constants. We calculate the energy for each set and then, sum them with different weights. In this work we use 20 replica to get smooth values between $-1$ and $+1$ for $J$.}.

Having the energy landscape, we use a painting algorithm to find minima. First we sort the energies from low to high. We start from the bottom of the list and paint the first element with a color. Then, we look at the next lowest energy state. If this state is a neighbor of the former state (It is different in only one spin), we paint it with the same color. If it is not a neighbor of any painted state, we associate a new color to that. As we go up the list, if we meet a state none of whose neighbors were painted before, it is a minimum. But if it has painted neighbors, then we paint it with the same color so it will belong to the same basin. In the case it has some neighbors which have been already painted with different colors, this indicates a border. In this method, members of any basin are painted with the same color and the border of colors show the transit states.

Basins are characterized by the minimum energy, their size and their depth. We already know the minimum energy of the basins. The size of a basin is the number of states which belongs to that basin (The number of states with the same color). Defining the depth of a basin is more tricky. Each basin has a border where the states over it might not have equal energies. Therefore, knowing all of the states on the border, we have three choices to use in defining the depth of a basin: the average energy on the border, the highest energy barrier and the lowest energy barrier. Subtracting the minimum energy from these values gives us mean depth, highest depth and the lowest depth, respectively. 





The connectivity graph does not give us any information about the energy barriers; therefore, we use the information of energy barriers to construct the {\em disconnectivity graph}\cite{1997JChPh.106.1495B}. The disconnectivity graph is generated in the following way. Imagine that the energy landscape is a real landscape consisting of mountains and valleys. Below this landscape, there is an underground sea. We start increasing the level of underground water. When the water level reaches to the deepest part of the landscape, a lake start to form. As we increase the level of underground water, more lakes (super basins) appear, and some times, lakes merge and give rise to larger lakes. Now in our graph, each node coresponds to a lake (super basin), and the links connect each lake to its ancestors --lakes that created it by merging. In a given water level a new lake can appear or a lake can shape from merging of other lakes; therefore, we associate a water level (energy) to each node. It shows the water level (energy) in which the lake (super basin) is created.

Spin glasses are well known for frustration, so one can ask if frustration has any impact on the network topology. We investigate this question by using coupling constants distributed between $a-1$ and $a+1$ where $a$ is constant. Here $a=0$ means our frustrated spin glass model and $a = -1$ represents a ferromagnetic Ising model. Therefore, the parameter $a$ controls the amount of frustration. We shall show that altering $a$ changes the shape of the degree distribution. 

We are also interested in various statistical properties akin to that of the degree distribution of the underlying graphs, the histograms of minimum energies, basin sizes and depths. To have good statistics for these PDFs , we ran the enumeration for different realizations of quenched random coupling constant and then we averaged the results. Our results will be demonstrated in the next section.

\section{Results and discussion}
\label{sec:res}
Looking at the probability density function of the energy of local minima (FIG. \ref{fig:comparison}), shows that it has a distorted Gaussian shape that follows Eq.~(\ref{eq:p_e}). Such a PDF of minima is also seen in the case of LJ clusters\cite{2000JPCM...12.6535H, 1999PhRvL..83.3214S, 2002JChPh.116.3777D}.

\begin{figure}
    \centering
    \includegraphics[width=\imagewidth]{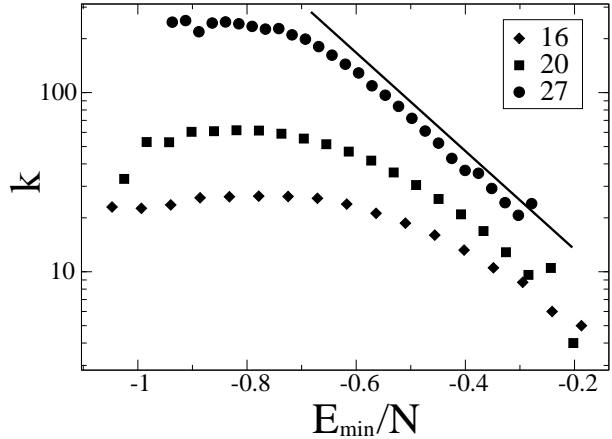}
    \caption{Degrees vs Energy of minima. It says that for large system we can consider the degree of a minimum as an exponential function of its energy: $k \sim e^{E_{min}}$.}
    \label{fig:K_E}
\end{figure}

Moreover, we investigate a relation between degrees of basins and their minimum energy (FIG.~\ref{fig:K_E}). 
As one can see $\ln k \sim -\frac{E}{J}$ relates the degree to energy. This is expected according to the argument in the analytical section of this article. But for minima with energies close to the global minimum, this relation does not hold any more. This might be due to the finite size effect, as it is more visible in the smaller systems.
\begin{figure}
    \centering
    \includegraphics[width=\imagewidth]{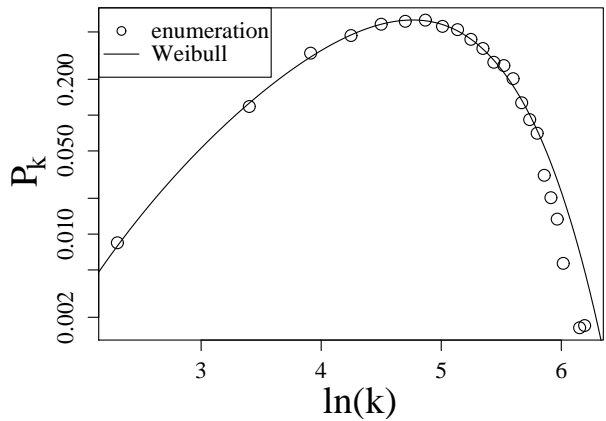}
    \caption{Probability density function of nodes' degree. Disks are the result of enumeration of 27 spins, averaged over 150 repeats. Solid line is Weibull function.}
    \label{fig:pdf_degree}
\end{figure}

Histogram of the numbers of the neighbors of a basin is the next quantity to study (FIG.~\ref{fig:pdf_degree}). It is fitted to log-Weibull distribution, meaning that the logarithms of $k$ follows Weibull distribution.
The Weibull distribution\cite{weibull} is a continuous probability distribution with the probability density function
\[
f(x;\kappa,\lambda) = \left \{ \begin{array}{ll}
{\kappa \over \lambda} \left({x \over \lambda}\right)^{\kappa-1} e^{-(x/\lambda)^\kappa} &  \mbox{if $x \geq 0$} \\
0 & \mbox{otherwise}\\ \end{array} \right. , \]
where $\kappa > 0$ is the shape parameter and $\lambda > 0$ is the scale parameter of the distribution. The best fitted values in here are  $\kappa=8.22$ $\lambda=4.84$.
Histograms of minimum depths have also an exponential PDF (FIG.~\ref{fig:pdf_min_depth}) indicating many shallow basins (short energy barriers) but very few deep basins (tall energy barriers).


We obtained a log-Weibull behavior for the distribution of degrees for our spin system (FIG.~\ref{fig:pdf_degree}) and a normal distribution for energy of minima (FIG.~\ref{fig:comparison}).The PDF of minimum energies is scaled with the number of spins. For all system sizes, we have a peak at about $0.5~J$. Thus, for any system size, the distribution peak is considerably far from origin. Assuming that $k \sim e^{-E_{min}}$, a log-normal behavior is expected for the PDF of degrees. This is  in consonant with FIG.~3 of reference \cite{amaral2000csw}, for short chains of random polymers and  with FIG.~9 of reference \cite{PhysRevE.73.036110}.
There are also contrasting reports of scale-free behavior for the topology of the energy landscape in some other systems\cite{doye_PhysRevLett.88.238701,rao2004pfn}.

It is important to know why there is such a difference. We note that these two systems --spin-glass and LJ clusters-- are both NP problems from the computational complexity point of view and at the same time, they are two standard models to study complex systems.

There are three main reasons for having different results for spin-glasses on one hand and for proteins and cluster of LJ particle on the other hand: a spin-glass is a frustrated system whereas the others are not. The energy density of different systems might be different from spin-glass and their data are to be interpreted in a different way.




\begin{figure}
    \centering
    \includegraphics[width=\imagewidth]{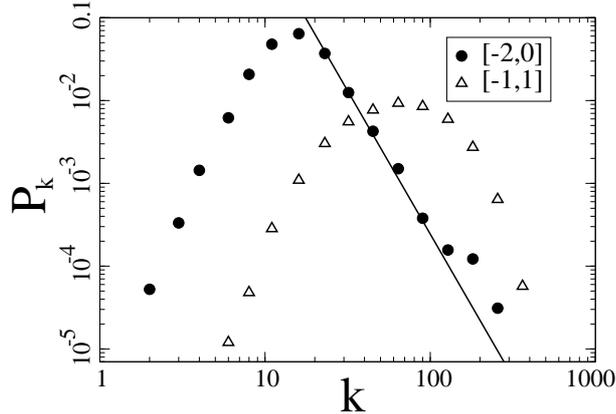}
    \caption{Frustration affects the probability density functions of degrees in the energy landscape networks. The results are for a 25-spin system. Disks shows the PDF for a model without frustration and triangles present PDF of the  frustrated spin-glass  model. The solid line has a slope of $-3.46$. This result suggest that absence of frustration may lead to something close to a power-law tail.}
    \label{fig:frust}
\end{figure}
\begin{figure}
    \centering

    \includegraphics[width=\imagewidth]{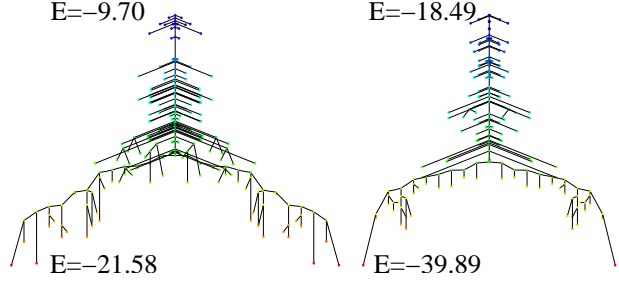}

    \caption{(color online) The disconnectivity graph of frustrated spin-glass is on the left and the one of unfrustrated spin-glass is on the right. The vertical direction represents the energy of the super basins, in a way that higher nodes have larger energies. The highest and the lowest energies are annotated. For a better understanding, we also showed the energies in colors: red shows the lowest energy and blue represents the highest energy. Here, the system size is 25 spins, and the coupling constant of the unfrustrated spin-glass is produced by shifting the coupling constants of the frustrated model by $+1$.}
    \label{fig:dis}
\end{figure}

\begin{figure}
   \centering
   \includegraphics[width=\imagewidth]{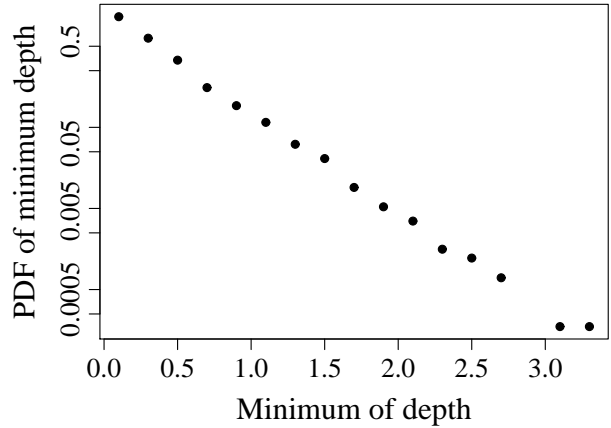}
   \caption{In this figure the PDF of minimum depths is demonstrated. Minimum depth is the difference between the minimum energy of a basin and the lowest energy on the border of that basin. It has an exponential distribution.}
\label{fig:pdf_min_depth}
\end{figure}

 
In order to remove  frustration from our model, it is enough to use a set of random coupling constant $J \in [-2,0]$. This is the minimal change to our model that abolishes frustration, while keeping the quenched randomness. The PDF of degree distribution of representing energy landscape network of this modified system is shown in FIG.~\ref{fig:frust}. One can see that this log-log plot of the PDF become similar to $y= y_0 - b|x-x_0|$, in which the slope of the power-law tail is $-3.46$. We already know that log-log plot of the same PDF for the case of frustrated spin-glass is close to a parabola. Interestingly, the PDF in the absence of frustration is very similar to the case of proteins\cite{rao2004pfn}. In addition to the PDF of the connectivity graph, we can look at the disconnectivity graph to compare the frustrated and none-frustrated models. In FIG.~\ref{fig:dis} one can see the disconnectivity graphs of a frustrated and unfrustrated system of 25 spins. The graph of frustrated system has more branches, which means the related landscape is rougher.


Reminding the mean field results, the log-Weibull shape of the histograms of the degrees is closely related to the Gaussian behavior of PDF of the energy minima. As it is shown in FIG.~\ref{fig:K_E}, the absolute value of minimum energy of a basin is proportional to the logarithm of the number of its neighboring basins. Then correlation between PDFs of degrees and minimum energies is a consequence of this logarithmic relation. Thus a power law distribution for degrees\cite{doye_PhysRevLett.88.238701} is the result of an exponential distribution of energy minima. But this possibility is ruled out because also in LJ clusters, energy minima follows a Gaussian distribution\cite{2006cond.mat.12205M, 2000JPCM...12.6535H}.




Another reason of the unlikeliness refers to different ways of interpreting the results. Usually we have few samples to calculate the probability density function directly. One way to overcome this problem is to use the cumulative distribution function (CDF) instead of the PDF, which gives us a plot with less noises. On the other hand, when we have data spanning only over a few orders of magnitude, by looking at the cumulative distributions, we can hardly distinguish between log-normal, log-Weibull and power-law distributions.

Using a spin glass model, we have the ability to run the enumeration on different quenched coupling constants which provides us with a rich statistics. Therefore, we can look at the average distribution over many realizations of the system instead of their cumulative distributions to dominate the noises. In this way, it is much easier to distinguish between these two types of the functions of power-law or log-Weibull. Our results are averaged over 150 realizations of 27-spin system  and 750 when  we had 25 spins.

\begin{figure}
 \centering
 \includegraphics[width=\imagewidth]{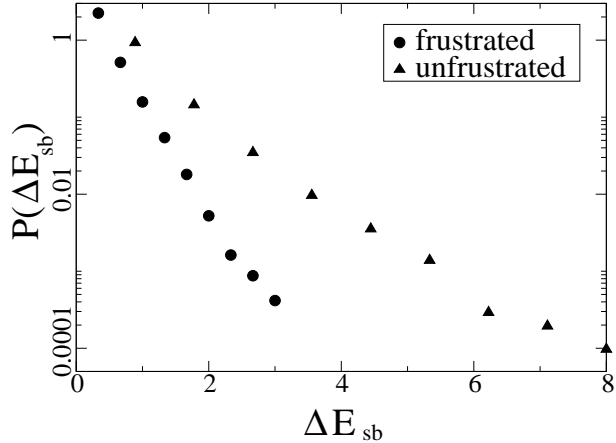}
 \caption{This figure shows the PDF of $\Delta E_{sb}$ which is the absolute value of the difference in the energy of two connected nodes in the disconnectivity graph. This result agrees with FIG.~\ref{fig:pdf_min_depth}. These curves are averaged over 200 realizations of 25 spins.}
 \label{fig:pdf_links_energy}
\end{figure}

One of our findings is the different behaviors of the minimum energies on the one hand and the minimum depth of the related basins on the other hand. As we mentioned before, minimum energies follow almost a Gaussian distribution whereas the minimum depths of the related basins follow an exponentially decaying distribution (FIG.~\ref{fig:pdf_min_depth}).
This fact suggests that we face a rough landscape. It means that we cannot define a large and smooth basin for the minima. Indeed, they have a rough basin that is full of many shallow basins. Any of these small basins  is  asymmetric in a way that the difference between its minimum depth and maximum depth is considerable. This is in agreement with our general perspective of spin-glass systems. The same conclusion is reached via disconnectivity graphs FIG.~\ref{fig:dis} and FIG.~\ref{fig:pdf_links_energy}.

\section{Conclusion}
\label{sec:con}
In this work we have shown that in the case of spin-glasses, the distribution of the energy minima is Gaussian and the associated network to the energy landscape has a log-Weibull degree distribution. This is verified by both numerical and analytical results. The result is in contrast with the case of the Lennard-Jones clusters which has a scale-free topology. This shows that the graph topology of an energy landscape is not universal and is highly related to  physical properties of the model system. Our results show that  the energy density function and frustration which are the most distinct differences between the two systems of spin-glasses and Lennard-Jones clusters are the most important factors. 


Frustration alters the probability density function of degrees in energy landscape network and changes the shape of the density function. In the absence of frustration, the probability density function has a tail that can be perceived as a power-law\footnote{one should be aware of the fact that the results span over only an order of magnitude}. This behavior of unfrustrated spinglass is similar to the behavior of proteins and Lennard-Jones clusters; therefore, one can consider the unfrustrated spin-glass as a better alternative to the frustrated spin-glass in the study of such systems.



\begin{acknowledgments}
We would like to thank Paolo De Los Rios for suggesting the significance of frustration, Marcel Ausloos for bringing our attention to the Weibull-distribution, Qasem Exirifard for careful reading the manuscript and Nima Hamedani Radja for valuable discussions.
\end{acknowledgments}




\bibliographystyle{unsrt}
\bibliography{spin_glass_network_hamid}

\end{document}